\magnification\magstephalf


\vskip 4cm
\centerline
{\bf On the Sugawara Current Algebra Proposal for M-Theory}
\centerline{}
\vskip 1cm
\centerline{Keith Glennon}
\vskip 0.5cm
\centerline{{\it Okinawa Institute for Science and Technology,}}
\centerline{{\it 1919-1 Tancha, Onna-son, Okinawa 904-0495, Japan}}
\centerline{keith.glennon@oist.jp\iffalse, ---, --- \fi} %
\vskip 2cm
\leftline{\sl Abstract}  

We examine the proposal of [29] that M-theory may admit a Sugawara-type current algebra formulation based on $E_{11} \otimes_s l_1$. Motivated by the role of generalized coordinates in E-theory, we ask whether current algebra relations of this type can be derived in a setting that includes those coordinates systematically. We show that such a construction can indeed be carried out for a rigid $E_{11}$ model in which the generalized coordinates are treated as inert under the rigid symmetry, in contrast with E-theory. We also argue that the bilinear form entering the Schwinger term requires closer scrutiny, since any natural ad-invariant extension of the $E_{11}$ Cartan-Killing form to $E_{11} \otimes_s l_1$ is degenerate.
\par 
\vskip2cm
\noindent

\vskip .5cm

\vfill
\eject

\medskip
{{\bf 1. Introduction}}
\medskip
\par 
\medskip
{{\bf 1.1 E-Theory Introduction}}
\medskip
\par
`E-theory' [1,2] proposes that M-theory [3,4] possesses a hidden $E_{11}$ Kac-Moody symmetry at low energies. It is argued that $E_{11}$ provides an eleven-dimensional explanation for the hidden $E_{11-D}$ symmetries [5,6] that arise when eleven-dimensional supergravity is dimensionally reduced to $D$ dimensions. It has been argued [7,8] that, upon restriction to the appropriate low-level field content, the dynamics of eleven-dimensional supergravity arise from a nonlinear realization [9,10] based on the coset 
$$
E_{11} \otimes_s l_1/I_c(E_{11}) \ \ .  \eqno(1.1.1)$$
Here $\otimes_s$ is the semi-direct product, $l_1$ is the vector representation of $E_{11}$, and $I_c(E_{11})$ is an involution-invariant local subgroup of $E_{11}$ that contains the Lorentz group ${\rm SO}(1,10)$. These objects will be defined quantitatively in Section 2. 
\par 
A key consequence of this reformulation is that eleven-dimensional M-theory and the low-energy limits of the IIA and IIB superstrings arise as different group-theoretic decompositions of $E_{11}$ [11]. In addition, all maximally gauged supergravities in all dimensions $D \geq 3$  are similarly described as decompositions of $E_{11}$ [12]. We thus obtain a group-theoretical perspective on some of the dualities of M-theory at low energies.
\par 
The trade-off for this reformulation is that one must introduce an infinite collection of fields. Many can be interpreted, at least in part, as electric-magnetic duals of the graviton and three-form of eleven-dimensional supergravity, while others have more subtle interpretations discussed in [2,13]. In addition we must extend our notion of spacetime [14] to include coordinates associated with each generator of the $l_1$ vector representation. At low levels these include coordinates associated with $P_{\mu}$, $Z^{\mu \nu}$, and $Z^{\mu_1 .. \mu_5}$, together with an infinite tower of further coordinates whose interpretation is less direct [2]. 
\par 
The $E_{11}$ program has been reviewed in [2,15,16] and extends beyond M-theory: it has been applied to four-dimensional gravity in [17] via the Kac-Moody algebra $A_1^{+++}$, and to the 26-dimensional closed bosonic string in [18] via the Kac-Moody algebra $K_{27}$, with implications for `bosonic M-theory' [19,20].
\par 
It will be useful to keep in mind that, as we review in Section 2, the coset model (1.1.1) contains at low levels the generators $K^a{}_b$, $P_{\mu}$, and $J_{ab}$ of
$$
{\rm GL}(11) \otimes_s \{P_{\mu}\}/{\rm SO}(1,10) \eqno(1.1.2)$$
Einstein's gravity can be described as a nonlinear realization based on this coset [15]. We will briefly set up and use this example in Section 4 to illustrate one of the difficulties we will encounter in a more familiar context.
\par
\vskip 0.5cm
\medskip
{{\bf 1.2 Current Algebra Introduction}}
\medskip
The development of current algebra in the 1960s was a major step on the path toward the Standard Model [21,22]. This led to proposals to reformulate entire physical theories in terms of currents as the basic variables [23]. It is in this context that the Sugawara current algebra was proposed [24,25]. This model proposed an abstract current algebra for vector and axial currents transforming in four dimensions under a rigid internal ${\rm SU}(3)$ symmetry group; when restricted to the vector currents, the algebra generalizes to any compact Lie algebra [32]. Concrete models of the Sugawara current algebra were soon found in [26] and [27].
\par
The historical development of particle physics suggests a useful analogy for the present problem. Current algebras were among the major advances beyond the conceptual framework of pure quantum electrodynamics [21,22], while internal symmetry and spontaneous symmetry breaking, expressed through nonlinear realizations, later became central ingredients of the Standard Model [28]. Taken together, current algebras, symmetry, and spontaneous symmetry breaking were key ideas in the transition from quantum electrodynamics to the Standard Model [28]. E-theory already contains two of these ingredients: a large symmetry algebra, $E_{11}$, and a nonlinear realization built from it. This naturally raises the question of whether a current algebra formulation also exists, and whether it can illuminate aspects of M-theory that are not manifest in the usual E-theory formulation.
\par 
\vskip 0.5cm
\medskip
{{\bf 1.3 The $E_{11}$ Current Algebra Proposal of [29]}}
\medskip
A proposal along precisely these lines was made in [29],
which put forward a `current algebra' formulation of M-theory based on\footnote{$^1$}{\sevenrm [29] includes additional spinor contributions, which we do not consider; we do not expect them to affect our discussion.}  $E_{11} \otimes_s l_1$. The proposal assumes that a current $J_{\mu}^{\alpha}$ can be associated with each generator of $E_{11} \otimes_s l_1$, and that these currents satisfy the `Sugawara commutation relations':
$$
[J_0^{\underline\alpha}(x),J_0^{\underline\beta}(y)] = - i \delta(x-y) f^{\underline \alpha \, \underline \beta}{}_{\underline \gamma} J_0^{\underline \gamma}(x)\eqno(1.3.1)$$
$$
[J_0^{\underline \alpha}(x),J_i^{\underline \beta}(y)] = - i \delta(x-y) f^{\underline \alpha \, \underline \beta}{}_{\underline \gamma} J_i^{\underline\gamma}(x) - i C \delta^{\underline\alpha \underline\beta} \partial_{i,x} \delta(x-y) \eqno(1.3.2)$$
In addition, the proposal assumes a Sugawara energy-momentum tensor bilinear in the currents, satisfying the commutation relations\footnote{$^2$}{\sevenrm (1.3.4) follows from (1.3.1) --- (1.3.3) [32] so we will not consider it further.}
$$
T_{\mu \nu} = {1 \over 2 C} {\cal C}^{\alpha \beta} [J_{\mu,\alpha} J_{\nu,\beta} + J_{\nu,\alpha} J_{\mu,\beta} - {1 \over 2} \eta_{\mu \nu} \eta^{\rho \sigma} J_{\rho,\alpha} J_{\sigma,\beta}] \eqno(1.3.3)$$
$$
[T_{00}(\vec{x}),T_{00}(\vec{y})] = (-i) \{ T_{0 \mu}(\vec{x}) + T_{0 \mu}(\vec{y}) \} \partial_{\mu} \delta(\vec{x} - \vec{y}) \eqno(1.3.4) $$
Here $f^{\underline\alpha \, \underline\beta}{}_{\underline\gamma}$ are taken as the structure constants of $E_{11} \otimes_s l_1$. In (1.3.1) --- (1.3.4), the coordinates and delta functions are those of ordinary eleven-dimensional spacetime; the generalized coordinates of E-theory do not appear. We will discuss the meaning of $\delta^{\underline\alpha \underline\beta}$ in the Schwinger term of (1.3.2) below.
\par
\vskip 0.5cm
\medskip
{{\bf 1.4 This Work}}
\medskip
The proposal of [29] is conceptually striking: it suggests that M-theory may admit a current-algebra formulation based on $E_{11} \otimes_s l_1$, together with quantum equations of motion\footnote{$^3$}{\sevenrm See equation (1.9) and (1.10) of [29]. We will not discuss these quantum equations of motion.} for the associated currents $J_{\mu}^{\underline{\alpha}}$. In that proposal, the quantum dynamics of M-theory are argued to be constrained directly by the symmetry structure of the low-energy theory. Related applications to branes and cosmology were explored in [30,31]. We therefore interpret [29] as a proposal to quantize M-theory starting from the symmetry structure of its low-energy theory.
\par 
The aim of the present work is thus to study this proposal. To do so, we focus on two structural aspects of (1.3.1) --- (1.3.3): the treatment of the generalized coordinates, and the bilinear form underlying the Schwinger term.
\par 
Since the relations (1.3.1) --- (1.3.3) do not incorporate the higher generalized coordinates that are central to E-theory, our first question is whether a corresponding generalized current algebra can be derived in a setting that includes those coordinates. In Section 3 we will show that such a derivation is indeed possible, in a framework in which the generalized coordinates are treated as inert under the rigid symmetry. To do this we will adapt the methods of [32] in a way that incorporates the higher coordinates systematically. This corresponds to a Type II nonlinear realization [2], in which the coordinates are treated as external parameters that are inert under the rigid symmetry. By contrast, as reviewed in Section 2.2, E-theory is based on a Type III nonlinear realization, in which the generalized coordinates transform non-trivially under rigid $E_{11}$.
\par
A second issue is that [29] studies $E_{11} \otimes_s l_1$; the Schwinger term and Sugawara energy-momentum tensor appear to require a non-degenerate ad-invariant bilinear form on this algebra. In Section 2.4 we will argue that, because $l_1$ is abelian, the natural ad-invariant extension of the $E_{11}$ bilinear form to $E_{11} \otimes_s l_1$ is necessarily degenerate. This suggests that the role of the bilinear form in the proposal of [29] requires closer examination. For this reason, the derivation given in Section 3 will construct an analogue of (1.3.1) --- (1.3.3) for $E_{11}$, rather than for $E_{11} \otimes_s l_1$, while incorporating the higher generalized coordinates.
\par
Our strategy is therefore the following. In Section 2 we first review the aspects of E-theory relevant to our discussion. In Section 3 we then derive a version of (1.3.1) --- (1.3.3) for a model with rigid internal $E_{11}$, incorporating all higher-level generalized coordinates and a non-degenerate bilinear form. In Section 4 we compare this derivation with the structure of E-theory and with the proposal of [29], discuss the obstacles in extending the construction to a Type III nonlinear realization and to a degenerate bilinear form, and draw conclusions.

\par 
\vskip 0.5cm
\medskip
{\bf 2. Review of $E_{11} \otimes_s l_1/I_c(E_{11})$}
\medskip
We now summarize the main aspects of E-theory that will be relevant to our current algebra discussion. We will first define the different parts of $E_{11} \otimes_s l_1/I_c(E_{11})$, and then sketch how the dynamics of M-theory at low energies are found from this setup.
\par
\vskip 0.5cm
\medskip
{\bf 2.1 Defining $E_{11}$, $l_1$, $I_c(E_{11})$}
\medskip
$E_{11}$ is an infinite-dimensional (Kac-Moody) Lie algebra defined by a Dynkin diagram
$$
\matrix{ & & & & & & & & & & & & & &  \oplus & 11 & & \cr
& & & & & & & & & & & & & &  | & & & \cr
\bullet & - & \bullet & - & \bullet & - & \bullet & - & \bullet & - & \bullet & - & \bullet & - & \bullet & - & \bullet & - & \bullet \cr
1 & &  2 & & 3 & &  4 & & 5 & & 6 & & 7 & & 8 & & 9 & & 10  }$$
Given this diagram, there is a well-defined procedure [1,2,15] by which $E_{11}$ can be decomposed in eleven dimensions with respect to node 11 as
$$
R^{\underline{\alpha}} \in E_{11} = \{ K^a{}_b \ ; R^{a_1 a_2 a_3} \ ; \ R^{a_1 .. a_6} \ ; \ R^{a_1 .. a_8,b} \ ; \ldots \ \ \ldots \} \eqno(2.1.1)$$
where $a=1,..,11$. The generators explicitly listed in (2.1.1) are at levels $0$, $1$, $2$, $3$. At levels $-1$, $-2$, $-3$ we also have $R_{a_1 a_2 a_3}$, $R_{a_1 .. a_6}$, $R_{a_1 .. a_8,b}$ not explicitly shown in (2.1.1). For the low-level generators displayed above, the level is given by the number of upper minus lower indices divided by 3. We denote a general generator of $E_{11}$ by $R^{\underline{\alpha}}$, where $\underline{\alpha}$ ranges over positive-root, negative-root, and Cartan generators; when no underline is used, $\alpha$ denotes the positive-root and Cartan sector.
\par 
We next associate the following coefficients
$$
{\cal A}_{\alpha} \in \{h_a{}^b;A_{a_1 a_2 a_3};A_{a_1..a_6};h_{a_1..a_8,b},...\} \eqno(2.1.2)$$
to the non-negative-level generators in (2.1.1), and later interpret them as the corresponding fields in the nonlinear realization of Section 2.2. The generator $K^a{}_b$ is at level zero and will lead to a field describing the graviton. $R^{a_1 a_2 a_3}$ is at level $+1$ and will lead to the three-form $A_{a_1a_2a_3}$ of M-theory. $R^{a_1 .. a_6}$ at level two will lead to its electric-magnetic dual $A_{a_1..a_6}$. $R^{a_1 .. a_8,b}$ at level three will lead to an electric-magnetic dual for the graviton which we will call the dual-graviton $h_{a_1..a_8,b}$. This is a mixed symmetry tensor $R^{a_1 .. a_8,b}$ which satisfies the irreducibility condition $R^{[a_1 .. a_8,b]} = 0$. The coset structure in (2.2.1) below means we will not associate fields to the negative root generators. The duality interpretation will be justified by equations (2.3.2) --- (2.3.3) below.
\par 
In the conventions of [14,15] the algebra is determined to be
$$
[K^a{}_b,K^c{}_d] = \delta^c{}_b K^a{}_d - \delta^a{}_d K^c{}_b \ , $$
$$
[K^a{}_b,R^{c_1 c_2 c_3}] = 3 \delta^{[c_1}{}_b R^{|a|c_2 c_3]} \ , \ [K^a{}_b,R_{c_1 c_2 c_3}] = - 3 \delta^a{}_{[c_1} R_{|b|c_2 c_3]} \ , \eqno(2.1.3) $$
$$
[R^{a_1 a_2 a_3},R_{b_1 b_2 b_3}] = 18 \delta^{[a_1 a_2}_{[b_1 b_2} K^{a_3]}{}_{b_3]} - 2 \delta^{a_1 a_2 a_3}_{b_1 b_2 b_3} D \ ,  \ [R^{a_1 a_2 a_3},R^{a_4 a_5 a_6}] = 2 R^{a_1 \ldots a_6} \ , \ \ldots $$
where $D:= \sum_{e=1}^{11} K^e{}_e$. These commutators determine the structure constants $f^{\underline{\alpha} \underline{\beta}}{}_{\underline{\gamma}}$ that will later enter the current algebra relations of Section 3.
\par 
The vector representation of $E_{11}$, denoted $l_1$, is generated by elements $l_A \in l_1$ where
$$
l_1 = \{P_a ; Z^{a_1 a_2} \ ; \ Z^{a_1 \ldots a_5} \ ; \ Z^{a_1 \ldots a_8} \ , \ Z^{a_1 \ldots a_7,b} \ ; \ \ldots \} \eqno(2.1.4)$$
Here $P_a$ generates the usual eleven-dimensional spacetime, while $Z^{a_1 a_2}$, $Z^{a_1\ldots a_5}$, and higher generators define additional directions in an infinite-dimensional generalized spacetime dictated by the representation theory of $E_{11}$. We take the $l_1$ representation to be abelian [14], transforming under $E_{11}$ as 
$$
[K^a{}_b,P_c] = - \delta^a{}_c P_b + {1 \over 2} \delta^a{}_b P_c \ , \ [R^{abc},P_d] = 3 \delta^{[a}_d Z^{bc]} \ ,  $$
$$
[R_{abc},P_d] = 0 \ , \ [R_{abc},Z^{de}] = - 6 \delta^{\,de}_{[ab} P_{c]} \ , \ \ldots \eqno(2.1.5)$$
\par 
We next introduce generalized coordinates associated with the generators $l_A$:
$$
x^{\Pi} \in \{x^{\mu},x_{\mu_1 \mu_2},x_{\mu_1\ldots\mu_5},\ldots\} \eqno(2.1.6)$$ 
These are interpreted as coordinates of a generalized spacetime generated by the $l_{\Pi}$, where $E_{11}$ acts via $[R^{\alpha},l_{\Pi}] = - (D^{\alpha})_{\Pi}{}^{\Sigma} l_{\Sigma}$. In Section 2.2 we will treat the ${\cal A}$ as fields dependent on the generalized coordinates $x^{\Pi}$. The proposal of [29] uses only the $x^{\mu}$ coordinates in the current algebra (1.3.1) --- (1.3.3).  
\par 
We now define an involution $I_c$ on $E_{11}$, $I_c^2 = I$, which satisfies $I_c(AB) = I_c(A) I_c(B)$. We take it to be given by
$$
I_c(K^a{}_b) = - \eta^{ad}   \eta_{bc} K^c{}_d \ , \ I_c(R^{a_1 a_2 a_3}) = - \eta^{a_1 b_1} \eta^{a_2 b_2} \eta^{a_3 b_3} R_{b_1 b_2 b_3} \ , $$
$$
I_c(R^{a_1 \ldots a_6}) = + \eta^{a_1 b_1} \ldots \eta^{a_6 b_6} R_{b_1 \ldots b_6} \ , \ I_c(R^{a_1 \ldots a_8,b}) = + \eta^{a_1 c_1} \ldots \eta^{a_8 b_8} \eta^{cd} R_{c_1 \ldots c_8,d} \ , \  \ldots \eqno(2.1.7)$$
Here $\eta_{ab}$ is taken to be the ${\rm SO}(1,10)$ Minkowski metric. We can now form involution invariant combinations
$$
J_{ab} = \eta_{ae} K^e{}_{b} - \eta_{be} K^e{}_a \ , \ S_{a_1a_2a_3} = R^{b_1 b_2 b_3} \eta_{b_1 a_1} \eta_{b_2 a_2} \eta_{b_3 a_3} - R_{a_1 a_2 a_3} \ , \ $$
$$
S_{a_1 \ldots a_6} = R^{b_1 \ldots b_6} \eta_{b_1 a_1} \ldots \eta_{b_6 a_6} + R_{a_1 \ldots a_6} \ , \ S_{a_1 \ldots a_8,b} = R^{c_1 \ldots c_8,d} \eta_{c_1 a_1} \ldots \eta_{c_8 a_8} \eta_{db} - R_{a_1 \ldots a_8,b} \ , \ \ldots \eqno(2.1.8)$$
Here we see that $I_c(E_{11})$ is a generalization of the Lorentz group ${\rm SO}(1,10)$.
\par
\vskip 0.5cm
\medskip
{\bf 2.2 The Nonlinear Realization of $E_{11} \otimes_s l_1/I_c(E_{11})$}
\medskip
We now discuss the nonlinear realization of 
$$
E_{11} \otimes_s l_1/I_c(E_{11}) \eqno(2.2.1)$$
In the nonlinear realization [2], the dynamics are described in terms of a local group element $g(x)$ of (2.2.1) subject to the transformation laws
$$
g \ \to \ g_0 g \ \ , \ \ g_0 \in E_{11} \ \ ; \ \ g \ \to \ gh \ \ , \ \ h \in I_c(E_{11}) \eqno(2.2.2)$$
Here $g_0$ is taken to be a rigid $E_{11}$ transformation, while $h$ is taken to be a local $I_c(E_{11})$ transformation. 
\par 
We take a group element of (2.2.1) to be parametrized as
$$
g(x) = g_l(x) g_E(x) \ \ , \ \ g_l(x) = \exp(x^{\Pi} l_{\Pi}) \ \ , \ \ g_E(x) = \Pi_{\alpha \geq 0} \exp({\cal A}_{\alpha}(x) R^{\alpha}) \ \ , \eqno(2.2.3)$$
In (2.2.3) we see the ${\cal A}$ are treated as fields depending on the generalized coordinates $x^{\Pi}$. The coset structure in (2.2.1) is essential because the local subgroup allows one to remove the fields associated with the negative roots. 
\par 
Under a rigid transformation (2.2.2), the group element (2.2.3) transforms as
$$
g(x) \to g_0 g(x) = [g_0 g_l(x) g_0^{-1}] g_0 g_E(x) = [g_l(x')] g_0 g_E(x) \ \ ; \ \ \ {\rm where}$$
$$
g_0 l_{\Pi} g_0^{-1} = D(g_0)_{\Pi}{}^{\Sigma} l_{\Sigma} \ \ , \ \ x'^{\Sigma} = x^{\Pi}  D(g_0^{-1})_{\Pi}{}^{\Sigma} \eqno(2.2.4)$$
This shows that the generalized spacetime coordinates transform under rigid $E_{11}$ transformations; they are therefore not inert. 
\par 
Because the coset in (2.2.1) includes both $E_{11}$ and a representation, $l_1$, in the form $E_{11} \otimes_s l_1$, it is a `Type III' nonlinear realization in the language of [2]. A `Type II' nonlinear realization is of the form $G/H$, where the fields in $g_G(x)$ depend on coordinates $x$ treated as external parameters unaffected by the group. The underlying spacetime coordinates do not transform as in (2.2.4); they simply remain inert. 
\par 
We now define a `generalized vielbein' through 
$$
g_E^{-1} l_{\Pi} g_E = E_{\Pi}{}^A l_A \eqno(2.2.5)$$
Under the transformations of (2.2.2) we see $E'_{\Pi}{}^A = D(g_0)_{\Pi}{}^{\Sigma} E_{\Sigma}{}^B D(h)_B{}^A$ so that $E_{\Pi}{}^A$ transforms as a vielbein on $E_{11}/I_c(E_{11})$. The components of the vielbein are given as an infinite-dimensional matrix
$$
E_{\Pi}{}^A = (\det e)^{-1/2} \left(\matrix{e_{\mu}{}^a & -3 e_{\mu}{}^c A_{c a_1 a_2} & \ldots \cr 0 & e_{[a_1}{}^{\mu_1} e_{a_2]}{}^{\mu_2} & \ldots \cr \vdots & \vdots & \ddots}\right) \ \ \ , \ \ \ e_{\mu}{}^a := (e^h)_{\mu}{}^a \ \ . \eqno(2.2.6)$$
This is reminiscent of double field theory [33,34], where the generalized vielbein similarly packages the spacetime vielbein together with additional fields such as the Kalb-Ramond field and the dilaton. The link to double field theory is made precise in [34,35].
\par 
We now construct the Maurer-Cartan form for this nonlinear realization as
$$
g^{-1} dg = dx^{\Pi} E_{\Pi}{}^A l_A + dx^{\Pi} G_{\Pi,\alpha} R^{\alpha} \eqno(2.2.7)$$

Using (2.2.5), we see from (2.2.7) that the vielbein $E_{\Pi}{}^A$ is associated to the generators $l_A$ of the abelian $l_1$ representation. By contrast, in the Type II construction considered in Section 3 there is no semidirect-product $l_1$ group factor, and hence no generalized vielbein arising from a relation of the form (2.2.5), even though the fields will still be allowed to depend on inert coordinates $x^{\Pi}$.
\par 
The Maurer-Cartan form is invariant under left-rigid $g(x) \to g_0 g(x)$ transformations. Under a right-local $g(x) \to g(x)h(x)$ transformation, the Maurer-Cartan transforms as a connection 
$$
g^{-1} d g \to h^{-1} [g^{-1} dg] h + h^{-1} d h. \eqno(2.2.8)$$
\par 
The $G_{\Pi,\alpha}$ coefficients in (2.2.7) may be interpreted as `covariant derivatives' of the fields ${\cal A}_{\alpha}$. By evaluating $[g_E^{-1} d g_E,l_{\Pi}]$ we find [36] $\partial_{\Pi} E_{\Sigma}{}^B = E_{\Sigma}{}^A G_{\Pi,A}{}^B = E_{\Sigma}{}^A  G_{\Pi,\alpha} (D^{\alpha})_A{}^B$. Thus $G_{\Pi,\alpha}$ are indeed components of the derivatives of the generalized vielbein. In Section 3 we will need the following abstract formula\footnote{$^4$}{\sevenrm If we set ${\scriptstyle g = \exp({\cal A}_{\alpha} R^{\alpha})}$ we would get ${\scriptstyle \exp({\cal A}_{\alpha} R^{\alpha}) d \exp({\cal A}_{\alpha} R^{\alpha}) = d {\cal A}_{\beta} [{1 - \exp(-{\cal A}_{\alpha} {\rm ad}_{R^{\alpha}}) \over {\cal A}_{\alpha} {\rm ad}_{R^{\alpha}}}]^{\beta}{}_{\alpha} R^{\alpha} = d {\cal A}_{\beta} e^{\beta}{}_{\alpha} R^{\alpha}}$ where we define ${\scriptstyle {\rm ad}_{R^{\alpha}}(R^{\beta}) := [R^{\alpha},R^{\beta}]}$. Our parametrization (2.2.3) gives a sum of products of this expression.} for the covariant derivatives [32,36] 
$$
G_{\Pi,\alpha} = e^{\beta}{}_{\alpha}({\cal A}) \partial_{\Pi} {\cal A}_{\beta} \eqno(2.2.9)$$
We emphasize that the partial derivatives in equation (2.2.9) arise in a linear fashion, which will become important when we construct a canonical momentum in Section 3. For example, the $A_{a_1 a_2 a_3}$ and $A_{a_1 .. a_6}$ covariant derivatives using the group element (2.2.3) are
$$
G_{\Pi,a_1 a_2 a_3} = e_{a_1 a_2 a_3}^{\mu_1 \mu_2 \mu_3} \partial_{\Pi} A_{\mu_1 \mu_2 \mu_3}, $$
$$
G_{\Pi,a_1 .. a_6} = e_{a_1 .. a_6}^{\mu_1 .. \mu_6}(\partial_{\Pi} A_{\mu_1 .. \mu_6} - A_{\mu_1 \mu_2 \mu_3} \partial_{\Pi} A_{\mu_4 \mu_5 \mu_6}) \ , \eqno(2.2.10)$$ 
These become the usual covariant derivatives of eleven-dimensional supergravity [2,7,8] when $\Pi = \nu$. Differentiating the matrix (2.2.6), one may re-express the result in terms of (2.2.10) and all other covariant derivatives of $E_{11}$ [2,7,8]. Although in our setup the roots $\alpha$ are non-negative in this formula, the formula still holds for positive and negative roots, a fact we will use in Section 3. 
\par
\vskip 0.5cm
\medskip
{\bf 2.3 Dynamics}
\medskip
To construct dynamics from the nonlinear realization, one requires objects which are covariant under the local $I_c(E_{11})$ transformations. The covariant derivatives $G_{\Pi,\alpha}$ transform, under local $I_c(E_{11})$ transformations, into linear combinations of other covariant derivatives. Combinations $G_{A,\alpha} = E_A{}^{\Pi} G_{\Pi,\alpha}$ such as $G_{a,b_1 b_2 b_3}$ will now also rotate on the $A$ indices under (2.2.5). For example, for an $I_c(E_{11})$ transformation generated by $\Lambda^{c_1 c_2 c_3}  S_{c_1 c_2 c_3}$, under (2.2.8) the field $G_{a,b_1 b_2 b_3}$ transforms to first order as [2,7,8,16]
$$
\delta G_{a,b_1 b_2 b_3} = 60 G_{a,b_1 b_2 b_3 e_1 e_2 e_3} \Lambda^{e_1 e_2 e_3} - 6 G_{a_1,(e[b_1)} \Lambda^e{}_{b_2 b_3]} - 3 \Lambda_{e_1 e_2 a} G^{e_1 e_2,}{}_{b_1 b_2 b_3} \ , $$
$$
\delta G^{a_1 a_2,}{}_{b_1 b_2 b_3} = 60 G_{a,b_1 b_2 b_3 e_1 e_2 e_3} \Lambda^{e_1 e_2 e_3} - 6 G_{a_1,(e[b_1)} \Lambda^e{}_{b_2 b_3]}  + 6 \Lambda^{a_1 a_2 c} G_{c,b_1 b_2 b_3} \eqno(2.3.1) $$
where we only list contributions from the first two levels $\Pi = a;a_1a_2$ of the $l_1$ representation and first three levels of $E_{11}$. In Section 4 we will emphasize that, because of the second line of (2.3.1), systematically incorporating the higher coordinates and local symmetries may lead to additional contributions absent from [29].
\par 
The fact that the covariant derivatives rotate into one another suggests that certain linear combinations of covariant derivatives may be preserved under local $I_c(E_{11})$ transformations. This suggests considering first order duality relations, such as [2,7,8,16]
$$
D_{a_1 a_2 a_3 a_4} = G_{[a_1,a_2 a_3 a_4]} - {1 \over 2 \cdot 4!} \epsilon_{a_1 .. a_4}{}^{b_1 .. b_7} G_{b_1,b_2 .. b_7} \eqno(2.3.2)$$
$$
D_{a,b_1 b_2} = (\det e)^{1/2} \omega_{a,b_1 b_2} - {1 \over 4} \epsilon_{b_1 b_2}{}^{c_1 .. c_9} G_{c_1,c_2 .. c_9,a} \eqno(2.3.3)$$
Here $(\det e)^{1/2} \omega_{a,b_1 b_2}$ is the spin connection of general relativity. Note that the duality relation in equation (2.3.3) is constructed from a connection. A connection is defined only up to gauge transformations, so this duality relation holds only modulo such gauge transformations. A proposed mechanism by which this can be treated as a full equation is discussed in [39].
\par 
When the effects of the higher coordinates are included, these first-order duality relations rotate into linear combinations of one another under local $I_c(E_{11})$ transformations, and it is consistent to set them to zero [7,8,2]. These first order duality relations thus encode the dynamics of the nonlinear realization, and are uniquely determined by the structure of $E_{11} \otimes_s l_1/I_c(E_{11})$. Second-order equations of motion for one of the two fields in a given duality relation can then be obtained in a well-defined manner, as discussed in [2,7,8,16]. In particular, one recovers the complete second-order equation for the three-form of eleven-dimensional supergravity, including the precise energy-momentum tensor [7,8,2].
\par
\vskip 0.5cm
\medskip
{\bf 2.4 Generalized Minkowski Metric $K^{\Pi \Sigma}$ and Cartan-Killing Metric ${\cal C}^{\alpha \beta}$}
\medskip
A generalization of the ${\rm SO}(1,10)$-invariant Minkowski metric $\eta^{\mu \nu}$ beyond level zero of $I_c(E_{11})$, to an $I_c(E_{11})$-invariant metric $K^{\Pi \Sigma}$ [40,41], is given by
$$
K_{\Pi \Sigma} = \left(\matrix{\eta_{\mu \nu} & 0 & & 0 & .. \cr 0 & 2! \eta^{\mu_1 \mu_2,\nu_1 \nu_2} & 0 & .. \cr 0 & 0 & 5! \eta^{\mu_1 .. \mu_5,\nu_1 .. \nu_5} & .. \cr \vdots & \vdots & \vdots & \ddots  }\right) \eqno(2.4.1)$$
\par 
In a symmetrizable Kac-Moody algebra [2], such as $E_{11}$, we can define a symmetric non-degenerate ad-invariant bilinear form ${\cal C}^{\underline\alpha \, \underline\beta} = {\cal C}(R^{\underline \alpha},R^{\underline \beta})$, which we refer to as the Cartan-Killing form, with inverse $({\cal C}^{-1})^{\underline \alpha \underline \beta}$. Ad-invariance of the Cartan-Killing metric means 
$$
{\cal C}([R^{\underline \alpha},R^{\underline \beta}],R^{\underline \gamma}) = {\cal C}(R^{\underline \alpha},[R^{\underline \beta},R^{\underline \gamma}]) \eqno(2.4.2)$$
that is, $f^{\underline \alpha \, \underline \beta}{}_{\underline \delta} {\cal C}^{\underline \delta \, \underline \gamma} = f^{\underline\beta \underline\gamma}{}_{\underline\delta} {\cal C}^{\underline\alpha \, \underline\delta}$, or equivalently $f^{\underline\beta(\underline\gamma}{}_{\underline\delta} {\cal C}^{|\underline\delta|\underline\alpha)} = 0$. In components, the Cartan-Killing form is given by [35]
$$
{\cal C}(K^a{}_b,K^c{}_d) = \delta^c{}_b \delta^a{}_d - {1 \over 2} \delta^a{}_b \delta^c{}_d \ \ , \ \ {\cal C}(R^{a_1 a_2 a_3},R_{b_1 b_2 b_3}) = 3! \delta^{a_1 a_2 a_3}_{b_1 b_2 b_3} \ \ , $$
$$
{\cal C}^{a_1 .. a_6}{}_{b_1 .. b_6} = - {1 \over 180} \delta^{a_1 .. a_6}_{b_1 .. b_6} \ \ , \ \ \ldots \eqno(2.4.3)$$
If one attempts to extend this ad-invariant bilinear form to include generators $l_A$ from the $l_1$ representation, then ad-invariance together with the abelian property of $l_1$ implies
$$
0 = {\cal C}(0,R^{\underline{\alpha}}) = (D^{\alpha})_B{}^C {\cal C}(l_A,l_C) \eqno(2.4.4)$$
giving degenerate directions. We thus see that the natural ad-invariant extension of the Cartan-Killing form to $E_{11} \otimes_s l_1$ is necessarily degenerate; the $l_1$ sector contains null directions.

\par 
\vskip 0.5cm
\medskip
{\bf 3. Current Algebras}
\medskip
\par
\medskip
{\bf 3.1 Historical Setup}
\medskip
We briefly sketch two historical examples in which the Sugawara current algebra emerges in a particular limit; for further details we refer to the original references. One of the first concrete examples of a Sugawara current algebra was presented in [25]. In this work it was argued that (1.3.1) --- (1.3.3) arise by taking a formal limit of a massive Yang-Mills theory given by
$$
{\cal L} = -{1 \over 4} F_{\mu\nu}^{a} F^{a \mu\nu} + {1 \over 2} m^2 A_{\mu}^{a} A^{a\mu} + g A_{\mu}^{a} J^{a\,\mu} \eqno(3.1.1)$$
where $F_{\mu\nu}^a = \partial_{\mu} A_{\nu}^a - \partial_{\nu} A_{\mu}^a + g f^{abc} A_{\mu}^b A_{\nu}^c$. The equations of motion are $D_{\nu} F^{\mu \nu a} + m^2 A^{\mu a} + g J^{\mu a} = 0$. The `formal limit' proposed is $m \to 0$, $g \to 0$, such that ${g \over m^2}$ is constant [25]. In this limit, [25] shows that the kinetic term in (3.1.1) is suppressed. The equations of motion then become $A^{\mu a} = - {g \over m^2} J^{\mu a}$, and so the Lagrangian is effectively of $JJ$ current-current form
$$
{\cal L} \approx - {g \over 2 m^2} J_{\mu}^a J^{\mu a} \eqno(3.1.2)$$
The energy-momentum tensor is thus of Sugawara form (1.3.3) in this limit. [25] shows that, while the original commutation relations and energy-momentum tensor associated with (3.1.1) are more complicated, relations of the form (1.3.1) --- (1.3.3) arise only after taking this limit. Here spacetime is treated as inert under a compact internal symmetry transformation in this model. 
\par 
Similar behavior occurs in the model of a rigid spontaneously broken ${\rm U}(1)$-invariant Lagrangian in [26]
$$
{\cal L} = - \partial_{\mu} S^{\dagger} \partial^{\mu} S + V(S^{\dagger} S) \ \ \ , \ \ \ S \to e^{i \alpha} S \ \ . \eqno(3.1.3)$$
Here we set $j_{\mu} = (-i) [S^{\dagger} \partial_{\mu} S - (\partial_{\mu} S^{\dagger}) S]$, and find $T_{\mu \nu} = 2 \partial_{(\mu} S^{\dagger} \partial_{\nu)} S + \eta_{\mu \nu} {\cal L}$. In polar variables $S = {1 \over \sqrt{2}} \rho e^{i \theta}$, the energy-momentum tensor and current algebra equal-time commutation relations are
$$
T_{\mu \nu} = \partial_{(\mu} \rho \partial_{\nu)} \rho - {1 \over 2} \eta_{\mu \nu} [\partial^{\sigma} \rho \partial_{\sigma} \rho - 2 V(\rho^2)] + {1 \over \rho^2} j_{(\mu} j_{\nu)} - {1 \over 2 \rho^2} \eta_{\mu \nu} j^{\sigma} j_{\sigma} \ \  , \ \ \eqno (3.1.4)$$
$$
[j_0(\vec{x},t),j_i(\vec{y},t)] = (-i) \partial_i [\rho^2(x) \delta(\vec{x} - \vec{y})] \ \ \ , \ \ \ j_{\mu} := \rho^2 \partial_{\mu} \theta  \eqno(3.1.5)$$
Only when the symmetry breaking constraint $\rho^2 = C$ is imposed do the energy-momentum tensor and current algebra take Sugawara form; in the same limit, the Lagrangian reduces to $JJ$ type.
\par 
\vskip 0.5cm
\medskip
{\bf 3.2  Maurer-Cartan Form of $E_{11}$ and its Rigid Transformations}
\medskip
We now begin the process of adapting the methods of [32] to derive a generalization of (1.3.1) --- (1.3.3) which incorporates the higher coordinates.
\par 
We begin by considering the case of a nonlinear realization with trivial subgroup $H$, so that the relevant group manifold is simply $E_{11}$, generated by $R^{\underline\alpha}$ as in (2.1.1). We let the fields ${\cal A}_{\underline \alpha}$ depend on the coordinates $x = x^{\Pi} $ of the $l_1$ representation, however here we treat the coordinates as inert under a rigid $E_{11}$ transformation (unlike in Section 2). Following [32], we are thus considering a Type II nonlinear realization, with fields now depending on all higher coordinates.
\par 
We take our group element of $E_{11}$ to be of the form
$$
g(x) = \exp({\cal A}_{\underline \alpha}(x) R^{\underline \alpha}) \eqno(3.2.1)$$
The Maurer-Cartan form for this group element is $g(x)^{-1} d g(x)$, where $d = dx^{\Pi} \partial_{\Pi}$ is a differential form on the $x^{\Pi}$; we find
$$
dx^{\Pi} (g^{-1} \partial_{\Pi} g) = dx^{\Pi} G_{\Pi,\underline \alpha} R^{\underline \alpha}  \ \ .  \eqno(3.2.2)$$
Under a left-rigid transformation $g(x) \to g_0 g(x) = g'(x)$, the Maurer-Cartan form is invariant, and hence the Cartan form coefficients are preserved, $G_{\Pi,\underline \alpha}' = G_{\Pi,\underline \alpha}$. Infinitesimally we set $g_0 = I + \Lambda$, with $\Lambda = \Lambda_{\underline \alpha} R^{\underline \alpha}$, so that $\delta g(x) = \Lambda g(x)$ and $\delta G_{\Pi,\underline \alpha} = 0$. The $\Pi$ indices are inert: unlike in Section 2.2, they do not transform under a rigid $E_{11}$ transformation.
\par 
A non-trivial transformation law arises by considering a right-rigid transformation $g(x) \to g(x) g_0 = g'(x)$. Infinitesimally, setting $g_0 = I + \Lambda$, with $\Lambda = \Lambda_{\underline \alpha} R^{\underline \alpha}$, so that $\delta g(x) = g(x) \Lambda$ or $g^{-1} \delta g = \Lambda$. In components, this reads as $(\delta {\cal A}_{\underline \beta}) e^{\underline \beta}{}_{\underline \alpha} R^{\underline \alpha} = \Lambda_{\underline \alpha} R^{\underline \alpha} $ so that
$$
\delta {\cal A}_{\underline \alpha} = \Lambda_{\underline \beta} (e^{-1})^{\underline \beta}{}_{\underline \alpha} \eqno(3.2.3)$$
The Maurer-Cartan form transforms by conjugation under a right-rigid transformation; we find the transformation $g^{-1} dg \to g_0^{-1} [g^{-1} d g] g_0$. The covariant derivatives transform infinitesimally as
$$
\delta G_{\Pi,\underline \alpha} = - \Lambda_{\underline \beta} f^{\underline{\beta} \, \underline{\gamma}}{}_{\underline \alpha} G_{\Pi,\underline \gamma} \eqno(3.2.4)$$
We can use the $e^{\beta}{}_{\alpha}({\cal A})$ to form a formal differential operator representation of $E_{11}$
$$
R^{\underline \beta} = (e^{-1})^{\underline \beta}{}_{\underline \alpha} \partial^{\underline \alpha} \ \ \ , \ \ \ \ \partial^{\underline \alpha} {\cal A}_{\underline \beta} = \delta^{\underline \alpha}{}_{\underline \beta} \ \ ; \ \ \ \partial^{\underline{\alpha}} := {\partial \over \partial {\cal A}_{\underline{\alpha}}} \ \ . \eqno(3.2.5)$$
We thus reproduce (3.2.3) via $\delta {\cal A}_{\underline \alpha} = (\Lambda_{\underline \beta} R^{\underline \beta}) {\cal A}_{\underline \alpha} = \Lambda_{\underline \beta} (e^{-1})^{\underline \beta}{}_{\underline \alpha}$. Note that $R^{\underline\alpha} = (e^{-1})^{\underline\alpha}{}_{\underline\rho} \partial^{\underline\rho}$ and $R^{\underline\beta} =  (e^{-1})^{\underline\beta}{}_{\underline\sigma} \partial^{\underline\sigma}$ imply
$$
\eqalign{[R^{\underline\alpha},R^{\underline\beta}]f &= f^{\underline\alpha \, \underline\beta}{}_{\underline\gamma} (e^{-1})^{\underline\gamma}{}_{\underline\sigma} \partial^{\underline\sigma} f
} \eqno(3.2.6) $$
Equivalently, one obtains $2 (e^{-1})^{[\underline\alpha}{}_{\underline\rho}  \partial^{|\underline\rho|} (e^{-1})^{\underline\beta]}{}_{\underline\sigma}   = f^{\underline\alpha \, \underline\beta}{}_{\underline\gamma} (e^{-1})^{\underline\gamma}{}_{\underline\sigma}$.
\par 
The integrability conditions $d \Omega = - \Omega \wedge \Omega $ for $\Omega = (e^{\beta}{}_{\alpha} d {\cal A}_{\beta} ) R^{\alpha}$ give similar relations. In components, this reduces to
$$
(\partial^{\alpha} e^{\beta}{}_{\gamma} - \partial^{\beta} e^{\alpha}{}_{\gamma} ) = -  (e^{\alpha}{}_{\rho} e^{\beta}{}_{\sigma} )  f^{\rho \sigma}{}_{\gamma} . \eqno(3.2.7)$$
This can also be written as $ (e^{-1}) ^{\rho}{}_{\alpha}  \partial^{\alpha} e^{\beta}{}_{\gamma} - (e^{-1})^{\rho}{}_{\alpha} \partial^{\beta} e^{\alpha}{}_{\gamma}  = -  (e^{\beta}{}_{\sigma} )  f^{\rho \sigma}{}_{\gamma}$.
\par 
\vskip 0.5cm
\medskip
{\bf 3.3 Quadratic Action Invariant Under Rigid $E_{11}$}
\medskip
We now introduce the simplest quadratic action compatible with these ingredients,
$$
S = \int d x {\cal L} \ \ \ ; \ \ \ dx := (\Pi_{\Pi} dx^{\Pi}) \eqno{(3.3.1)}$$
setting
$$
{\cal L} = {1 \over 2} C K^{\Pi \Sigma} G_{\Pi,\underline \alpha} G_{\Sigma,\underline \beta} {\cal C}^{\underline{\alpha} \, \underline{\beta}} \eqno{(3.3.2)}$$
Since this action integrates over an infinite-dimensional generalized spacetime, it should be understood formally [45,46]. In what follows we will mainly be concerned with the local Lagrangian density, and similar formal integrals have appeared previously in the E-theory literature. Using (3.2.2), the Lagrangian reads as
$$
{\cal L} = {1 \over 2} C K^{\Pi \Sigma}  (\partial_{\Pi} {\cal A}_{\underline \gamma}) (\partial_{\Sigma} {\cal A}_{\underline \delta}) (e^{\underline \gamma}{}_{\underline \alpha} {\cal C}^{\underline{\alpha} \,  \underline{\beta}} e^{\underline \delta}{}_{\underline \beta}) \eqno{(3.3.3)}$$
Here $C$ is a constant. The quadratic form appearing in the local Lagrangian density is well-defined algebraically because $K^{\Pi \Sigma}$ and ${\cal C}^{\underline{\alpha} \,  \underline{\beta}}$ are non-degenerate. Extending this model to $E_{11} \otimes_s l_1$ would force the natural ad-invariant Cartan-Killing form to be degenerate. For this reason we restrict attention here to $E_{11}$. 

\par 
\vskip 0.5cm
\medskip
{\bf 3.4 Rigid $E_{11}$ invariance of ${\cal L}$ and Current Densities}
\medskip
We now show the action (3.3.1) is invariant under the rigid $E_{11}$ transformations of Section 3.2. Under left-rigid transformations the Lagrangian is immediately invariant. For right-rigid transformations, $\delta {\cal L}$ vanishes due to the ad-invariance of ${\cal C}^{\underline{\alpha} \underline{\beta}}$ of (2.4.2):
$$
\eqalign{ \delta {\cal L} &= C K^{\Pi \Sigma} G_{\Pi,\underline\alpha} (\delta G_{\Sigma,\underline\beta}) {\cal C}^{\underline\alpha \underline\beta} \cr 
&= - C K^{\Pi \Sigma} \Lambda_{\underline\delta} f^{\underline\delta (\underline\gamma}{}_{\underline\beta} {\cal C}^{|\underline\beta| \underline\alpha)} G_{\Pi,\underline\alpha} G_{\Sigma,\underline\gamma} = 0 } \eqno{(3.4.1)}$$
\par 
We can thus define a conserved current [32]
$$
J^{\Pi,\underline\gamma} = {\delta {\cal L} \over \delta (\partial_{\Pi} {\cal A}_{\underline\beta})} {\delta {\cal A}_{\underline\beta} \over \delta \Lambda_{\underline\gamma}} \eqno(3.4.2)$$
For the specific Lagrangian of (3.3.2), this reads as
$$
J^{\Pi,\underline\gamma} = C K^{\Pi \Sigma} {\cal C}^{\underline\gamma \, \underline\beta} G_{\Sigma,\underline\beta}  \eqno(3.4.3) $$
We now set
$$
J_{\Sigma,\underline\alpha} = C G_{\Sigma,\underline \alpha} \eqno(3.4.4)$$
The Lagrangian can now be written in $JJ$ form
$$
{\cal L} = {1 \over 2 C} J_{\Pi,\underline \alpha} J^{\Pi,\underline \alpha} \ \ \  \eqno(3.4.5)$$
and the corresponding Sugawara energy-momentum tensor is
$$
T_{\Pi \Sigma} = {1 \over 2C} {\cal C}^{\underline \alpha \, \underline\beta} [ J_{\Pi,\underline \alpha} J_{\Sigma,\underline \beta} + J_{\Sigma,\underline \alpha} J_{\Pi,\underline \beta} - {1 \over 2} K_{\Pi \Sigma} (K^{\Omega \Lambda} J_{\Omega,\underline \alpha} J_{\Lambda,\underline \beta}) ] \eqno(3.4.6)$$
We have thus obtained the natural analogue of the Sugawara energy-momentum tensor (1.3.3) from the action (3.3.1).
\par
\vskip 0.5cm
\medskip
{\bf 3.5 Canonical Momenta and Current Algebra}
\medskip
We now face the question of how to define a canonical momentum. Such a choice breaks manifest (generalized) spacetime covariance; in the present generalized setting, this issue becomes even more acute. In an ordinary spacetime with coordinates $x^{\mu}$, one can immediately define $\pi^{\alpha}({\bf x}) = {\delta {\cal L} \over \delta (\partial_0 {\cal A}_{\alpha})}$, thereby singling out the time coordinate $x^0$. However, in a generalized spacetime one would have to define
$$
\pi^{\underline\gamma}(x^0,x^i;x^{\mu_1 \mu_2};x^{\mu_1\ldots\mu_5},\ldots) = {\delta {\cal L} \over \delta (\partial_0 {\cal A}_{\underline\gamma})} \eqno(3.5.1)$$
as the canonical momentum. This first isolates the level-zero coordinates, and then singles out the time component within that sector. We leave the question of a more democratic definition for the canonical momentum to later work. Here we take equation (3.5.1) as our definition. For the specific Lagrangian (3.3.2), the canonical momentum at fixed $x^0$ is
$$
\eqalign{\pi^{\underline\gamma}(x) &= C K^{0\Sigma} e^{\underline\gamma}{}_{\underline\alpha} {\cal C}^{\underline\alpha \, \underline\beta} G_{\Sigma,\underline\beta} } \eqno{(3.5.2)}$$ 
In what follows, our $x$ notation is a shorthand assuming equal times. From the definition (3.5.1), we can express the canonical momentum in terms of the current $J^{0,\underline\gamma}$ as
$$
J^{0,\underline\gamma} = {\delta {\cal L} \over \delta (\partial_0 {\cal A}_{\underline{\beta}})} {\delta {\cal A}_{\underline{\beta}} \over \delta \Lambda_{\underline\gamma}} = \pi^{\underline\beta} (e^{-1})^{\underline\gamma}{}_{\underline\beta} \eqno(3.5.3)$$
\par 
We now propose the equal-time commutation relations\footnote{$^5$}{\sevenrm Here ${\scriptstyle x}$ and ${\scriptstyle y}$ are vectors in the ${\scriptstyle l_1}$ representation which do not include the ${\scriptstyle x^0,y^0}$ coordinate. We suppress the time parameter in the fields to keep our notation simple. ${\scriptstyle \delta(x-y)}$ is formally an infinite product ${\scriptstyle \Pi_{\Sigma'} \delta(x^{\Sigma} - y^{\Sigma'})}$, where ${\scriptstyle \Sigma'}$ does not include the lowest level ${\scriptstyle 0}$ index.}
$$
[{\cal A}_{\underline\alpha}(x),\pi^{\underline\beta}(y)] = i \delta_{\underline\alpha}{}^{\underline\beta} \ \delta(x-y) \eqno(3.5.4)$$
Following [32], and using (3.5.3), (3.4.2)–(3.4.3), (3.2.7), and (3.5.4), one finds the equal-time commutators
$$
[J^{0,\underline\alpha}(x),J^{0,\underline\beta}(y)] =  - i  \delta(x-y)  f^{\underline\alpha \, \underline\beta}{}_{\underline\gamma} J^{0,\underline\gamma}(x) \eqno(3.5.5)$$
$$
[J^{0,\underline\alpha}(x),J^{\Pi',\underline\beta}(y)] = - i f^{\underline\alpha \, \underline\beta}{}_{\underline\gamma} J^{\Pi',\underline\gamma}(x) \delta(x-y) - i  (C K^{\Pi'\Sigma} {\cal C}^{\underline\alpha \, \underline\beta}) \partial_{\Sigma,x} \delta(x-y) \eqno(3.5.6) $$
Here $\Pi'$ denotes any generalized coordinate index other than the level-zero time index $0$. 
\par 
We have thus found a Sugawara-type current algebra analogous to that postulated in [29], now incorporating higher generalized coordinates. This, however, was based on assumptions characteristic of a Type II nonlinear realization, with the coordinates treated as inert under a rigid $E_{11}$ symmetry. The resulting current and commutation relations are given in (3.4.3), (3.5.5), and (3.5.6). We return to this point in Section 4.
\par 
We note that the derivation of (3.5.5) does not require the explicit form of the Lagrangian. There is a Lagrangian-independent relationship between the current $J^{0,\underline\gamma}$ and the canonical momentum $\pi^{\underline\gamma}$ in (3.5.3). The only quantities appearing on the right-hand side of (3.5.3) are the fields ${\cal A}_{\underline\alpha}$ and $\pi^{\underline\beta}$. We need only assume that the canonical momenta $\pi^{\underline\beta}$ arise from some Lagrangian, and that they satisfy the equal-time commutation relations (3.5.4). 
\par 
To evaluate (3.5.6), the actual form of the canonical momentum as derived from the specific Lagrangian in (3.3.2) is needed. The current is explicitly
$$
J^{\Pi',\underline\gamma} = {\delta {\cal L} \over \delta (\partial_{\Pi'} {\cal A}_{\underline\beta})} {\delta {\cal A}_{\underline\beta} \over \delta \Lambda_{\underline\gamma}} = (C K^{\Pi'\Sigma} {\cal C}^{\underline\gamma \, \underline\beta}) G_{\Sigma,\underline{\beta}} \eqno(3.5.7)$$
A different Lagrangian model could therefore produce no Schwinger term, or more general higher-derivative Schwinger terms [42,43].
\par
\vskip 0.5cm
\medskip
{\bf 4 Comparison to E-Theory}
\medskip
The proposal of [29] is conceptually striking and provides the main motivation for the present analysis. However, comparison with E-theory shows that the Sugawara-type current algebra derived in Section 3 relies on assumptions that differ in important respects from those of E-theory.
\par
The construction of Section 3 required a Type II nonlinear realization, rather than the Type III nonlinear realization used in E-theory. In addition, it did not use the coset structure of $E_{11} \otimes_s l_1/I_c(E_{11})$ or the local $I_c(E_{11})$ subgroup that plays a central role in the E-theory construction. Moreover, in the present derivation the higher coordinates enter only as inert generalized coordinates, whereas in E-theory they play an essential dynamical role. As illustrated by (2.3.1), higher-level derivatives mix with lower-level fields and are crucial in the derivation of the second-order equations of motion of eleven-dimensional supergravity [7,8,2]. By contrast, [29] appears to lead to an apparent violation of the equivalence principle discussed after equation (98), and to field equations in equation (84) that are reminiscent of, but not identical to, the Einstein equations. This leaves open the possibility that a formulation which systematically incorporates the higher coordinates and local symmetries of a Type III nonlinear realization would modify these features.
\par
This suggests that the difficulties encountered in [29] may not be incidental, but instead reflect a structural mismatch between the Sugawara-type current-algebra framework considered here and the Type III nonlinear realization underlying E-theory. In particular, [29] uses only the ordinary spacetime coordinates, whereas the E-theory dynamics depend essentially on the higher coordinates together with the local $I_c(E_{11})$ symmetry.
\par
A second issue concerns the bilinear form entering the Schwinger term. In Section 2.4 we argued that the natural ad-invariant extension of the $E_{11}$ Cartan-Killing form to $E_{11} \otimes_s l_1$ is degenerate. Equations (59.3) and (59.4) of [29] appear to rely on such a degenerate bilinear form, and these relations then enter the subsequent derivation of equation (84).
\par
By contrast, in Section 3 we derived (3.4.3), (3.5.6), and (3.5.7) using the non-degenerate ad-invariant bilinear form on $E_{11}$, namely its Cartan-Killing form. If the bilinear form implicitly used in [29] is not the standard ad-invariant one, then its definition and properties require separate justification; at present this point is not made explicit there.
\par
We now consider some of the difficulties that arise in attempting to extend the construction to a Type III nonlinear realization. A simple illustration is provided by the gravity coset formulation of [15]. If such Sugawara current-algebra techniques extended directly to models in which the spacetime coordinates themselves transform, it would be natural to ask whether an analogous reformulation already exists for general relativity in the coset formulation based on (1.1.2). For $D=11$, the Maurer-Cartan form of ${\rm GL}(11) \otimes_s \{P_{\mu}\}/{\rm SO}(1,10)$ is
$$
g^{-1} dg = dx^{\mu} e_{\mu}{}^a P_a + dx^{\mu} G_{\mu,a}{}^b K^a{}_b \ \ , \ \ 
e_{\mu}{}^a = (e^h)_{\mu}{}^a \ \ , \ \ 
G_{\mu,a}{}^b = (e^{-1} \partial_{\mu} e)_a{}^b \ \ , \eqno(4.1)
$$
where we take $g(x) = \exp(x^{\mu} P_{\mu}) \exp(h_a{}^b(x) K^a{}_b)$. The ${\rm SO}(1,10)$-invariant metric is the Minkowski metric $\eta^{ab}$. One may then form the object $g_{\mu \nu} = e_{\mu}{}^a \eta_{ab} e_{\nu}{}^b$, 
which is invariant under the local ${\rm SO}(1,10)$ subgroup. An action yielding second-order equations of motion for $g_{\mu \nu}$ is then of the form
$$
S = \int d^{11} x \sqrt{-g} \left[
{C_4 \over 4} (\partial^{\mu} \ln g)(\partial_{\mu} \ln g)
- {C_5 \over 2} (\partial^{\mu} \ln g)(\partial^{\nu} g_{\nu \mu})
+ {C_2 \over 4} (\partial^{\mu} g^{\nu \rho})(\partial_{\mu} g_{\nu \rho})
\right.
$$
$$
\left.
\phantom{S = \int d^{11} x \sqrt{-g} \bigl[}
- {C_1 \over 2} (\partial^{\mu} g^{\nu \rho})(\partial_{\nu} g_{\rho \mu})
+ C_3 g_{\mu \nu} \partial_{\rho} g^{\rho \mu} \partial_{\sigma} g^{\sigma \nu}
\right] \eqno(4.2)
$$
Requiring diffeomorphism invariance fixes the coefficients to be $C_1 = C_2 = C_4 = C_5 = 1$ and $C_3 = 0$ [15]. With these values, (4.2) reduces to the Einstein-Hilbert action up to a total derivative [15]. We are not aware of any set of assumptions that reduces the action (4.2) to a Sugawara-type $JJ$ form. Even if one formally defines currents from the quantities appearing in (4.2), it is not clear why they should satisfy a Sugawara-type current algebra. This is reminiscent of the examples reviewed in Section 3.1: for instance, the theory defined by (3.1.1) takes a Sugawara-type $JJ$ form only in the formal limit $m \to 0$, $g \to 0$. Something analogous may therefore be required for (4.2), if such a reformulation exists at all.
\par 
The present work clarifies the conditions under which a Sugawara-type current algebra can be derived for $E_{11}$, and thereby helps identify what a future current-algebra formulation of M-theory would need to incorporate in order to be compatible with E-theory. In this respect, the general relativity example discussed above serves as a useful test case for the difficulties that arise when one attempts to extend such a construction to a setting in which the spacetime coordinates themselves transform. A genuine current-algebra reformulation of general relativity would likely be a highly non-trivial undertaking, and may be viewed as a natural intermediate step toward any corresponding extension to $E_{11}$.
\par 
\vskip 0.5cm
\medskip
{\bf Acknowledgements}
\medskip
We would like to thank Peter West for discussions. We also wish to thank Hirotaka Sugawara for his hospitality at KEK during KEK-Theory 2025. This work was supported by the Quantum Gravity Unit at the Okinawa Institute of Science and Technology (OIST).
\par 
\rightskip=0pt plus 2em \spaceskip=0pt \xspaceskip=0pt \pretolerance=10000 \tolerance=10000
\medskip
{\bf References}
\medskip
\item{[1]} P. West, {\it $E_{11}$ and M Theory}, Class. Quant. Grav.  {\bf 18}, (2001) 4443, hep-th/ 0104081.
\item{[2]} P. West,{\it A brief review of E theory}, Proceedings of Abdus Salam's 90th  Birthday meeting, 25-28 January 2016, NTU, Singapore, {\bf Vol 31}, No 26 (2016) 1630043, \break arXiv:1609.06863.
\item{[3]} E. Witten, {\it String theory dynamics in various dimensions}, Nucl.\ Phys.\ {\bf B443} (1995) 85, arXiv:hep-th/9503124.
\item{[4]} M. J. Duff, {\it M-theory (The Theory Formerly Known as Strings)},  \break  Int.J.Mod.Phys.A11:5623-5641,1996, hep-th/9608117.
\item{[5]} E. Cremmer and B. Julia, {\it The N=8 Supergravity Theory. 1. The Lagrangian}, \break Phys.Lett.B80:48-51,1978.
\item{[6]} B. Julia, {\it Group Disintegrations},
in {\it Superspace and Supergravity}, eds. S. W. Hawking and M. Rocek,
Cambridge Univ. Press,1981, pp.331-350.
\item{[7]} A. Tumanov and P. West, {\it E11 must be a symmetry of strings and branes },  Phys. Lett. B 759 (2016) 663–671, arXiv:1512.01644 [hep-th]. 
\item{[8]} A. Tumanov and P. West, {\it E11 in 11D}, Phys.Lett. B758 (2016) 278, arXiv:1601.03974.
\item{[9]} S. R. Coleman, J. Wess and B. Zumino, {\it Structure of Phenomenological Lagrangians. I}, Phys.Rev.177:2239-2247,1969.
\item{[10]} C. G. Callan Jr., S. R. Coleman, J. Wess and B. Zumino, {\it Structure of Phenomenological Lagrangians. II}, Phys.Rev.177:2247-2250,1969.
\item{[11]} West, P. {\it The IIA, IIB and eleven dimensional theories and their common E11 origin}, Nucl. Phys. B693 (2004) 76-102, hep-th/0402140.
\item{[12]} F. Riccioni and P. West, {\it The $E_{11}$ origin of all maximal supergravities}, JHEP 0707 (2007) 063; arXiv:0705.0752.
\item{[13]} L. J. Romans, {\it Massive N = 2A supergravity in ten dimensions}, Phys. Lett. B 169 (1986) 374.
\item{[14]} P. West, {\it $E_{11}$, SL(32) and Central Charges}, Phys. Lett. {\bf B 575} (2003) 333-342,  hep-th/0307098.
\item{[15]} P. West, {\it Introduction to Strings and Branes}, Cambridge University Press, 2012.
\item{[16]} Glennon, K. {\it Applications of Kac-Moody Algebras to Gravity and String Theory.} \break (2023), arXiv:2312.11454.
\item{[17]} Glennon, K. and West, P. {\it Gravity, dual gravity
and $A_1^{+++}$. IJMPA, 35(14), (2020) p.2050068. arXiv:2004.03363
\item{[18]} Glennon, K. and West, P., 2024. {\it $K_{27}$ as a symmetry of closed bosonic strings and branes}. IJMPA, 40(2), p.2450155. arXiv:2409.08649
\item{[19]} Glennon, K. {\it Bosonic M-Theory From a Kac-Moody Perspective}, (2025), \break arXiv: 2501.03000.
\item{[20]} G. Horowitz and L. Susskind, {\it Bosonic M-theory}, J.Math.Phys.42:3152-3160,2001, hep-th/0012037.
\item{[21]} Pietschmann, H. {\it The early history of current Algebra.} The European Physical Journal H, 36(1), (2011) pp.75-84.
\item{[22]} Iliopoulos, J. {Fifty Years That Changed Our Physics}, Proceedings, 51st Rencontres de Moriond \break on Electroweak Interactions and Unified Theories : La Thuile, Italy, March 12-19, 2016, 15-30.
\item{[23]} R. F. Dashen and D. H. Sharp, {\it Currents as Coordinates for Hadrons}, \break Phys.Rev
Phys.Rev.Lett.18:1029-1032,1967..165:1857-1866,1968. D.J. Gross, {\it Nonrelativistic Quantum Mechanics Formulated in Terms of Currents}. Physical Review, 177(5), p.1843.
\item{[24]} H. Sugawara, {\it A Field Theory of Currents}, Phys.Rev.170:1659-1662,1968.
\item{[25]} C. M. Sommerfield, {\it Currents as Dynamical Variables}, Phys.Rev.176:2019-2025,1968.
\item{[26]} K. Bardakci, Y. Frishman and M. B. Halpern, {\it Structure and Extensions of a Theory of Currents}, Phys.Rev.170:1353-1359,1968.
\item{[27]} Y. Freundlich and D. Lurie, {\it Sugawara Model and Goldstone Bosons}, \break  Phys.Rev.D1:1660-1662,1970.
\item{[28]} M. Kobayashi, {\it Personal recollections on chiral symmetry breaking}, \break Prog.Theor.Exp.Phys. 2016 (2016) 07B101.
\item{[29]} H. Sugawara, {\it Current algebra formulation of M-theory based on $E_{11}$ Kac–Moody algebra},  IJMPA {\bf 32} (2017) 1750024, arXiv:1701.06894.
\item{[30]} S. Shiba and H. Sugawara, {\it M2- and M5-Branes in E11 Current Algebra Formulation of M-Theory},
Int.J.Mod.Phys.A33:1850051,2018, arXiv:1712.05123 [hep-th].
\item{[31]} S. S. Funai and H. Sugawara, {\it Current Algebra Formulation of Quantum Gravity and Its Application to Cosmology}. Prog.Theor.Exp.Phys.2020(9):093B08,2020. \break  arXiv:2004.02151.
\item{[32]} A. I. Solomon, {\it Nonlinear Realizations, the Sugawara Model and Goldstone Bosons}, Lett.Nuovo.Cim.4:337-340,1970.
\item{[33]} O. Hohm, C. Hull, and B. Zwiebach. {\it Generalized metric formulation of double field theory}. Journal of High Energy Physics, 2010(8), pp.1-35. arXiv:1006.4823.
\item{[34]} P. West. {\it E11, generalised space–time and IIA string theory}. Physics Letters B, 696(4), pp.403-409. arXiv:1009.2624
\item{[35]} P. West. {\it  Local symmetry and extended space–time}, IJMPA A 40 (2025) 24, 2550100 arXiv:2504.18229.
\item{[36]} A Tumanov, and P. West. {\it Generalised vielbeins and non-linear realisations}. JHEP 2014(10), pp.1-39. arXiv:1405.7894.
\item{[37]} I. Bandos, N. Berkovits, and D. Sorokin, D. {\it Duality-symmetric eleven-dimensional supergravity and its coupling to M-branes}. Nuclear Physics B, 522(1-2), pp.214-233. arXiv:9711055
\item{[38]} A.J. Nurmagambetov. {\it Hidden symmetries of M-theory and its dynamical realization}. SIGMA. Symmetry, Integrability and Geometry: Methods and Applications, 4, p.022. arXiv:0802.2638
\item{[39]} N. Boulanger, PP. Cook, J. O'Connor, J. and P. West, P. {\it Unfolding $E_{11}$}. SciPost Physics, 18(5), p.149. arXiv:2410.21206
\item{[40]} M. Pettit, and P. West. {\it An $E_{11}$ invariant gauge fixing}. IJMPA, 33(01), p.1850009. arXiv:1710.11024
\item{[41]} M. Pettit. {\it E theory: its algebra, gauge fixing, and 7D equations of motion}. (Doctoral dissertation, King's College London).
\item{[42]} P. A. M. Dirac, {\it The Conditions for a Quantum Field Theory to be Relativistic}, Rev.\ Mod.\ Phys.\ {\bf 34} (1962) 592.
\item{[43]} S. G. Brown, {\it Dirac-Schwinger Covariance Condition in Canonical Theories}, Phys.\ Rev.\ {\bf 158} (1967) 1608.
\item{[45]}  West, P. {\it Irreducible representations of E theory}. IJMPA, {\bf 34} (2019) 24, 1950133 arXiv:1905.07324
\item{[46]} Bossard, G, Boulanger, N., and O'Connor, J. {\it Higher dualities in E11 exceptional field theory}. arXiv:2602.22491

\end